\documentclass{iau} 
\usepackage{graphicx}
\newcommand{\rcor}{$R_{\rm cor}$}
\newcommand{\rbar}{$a_{\rm bar}$}
\newcommand{\omegabar}{$\Omega_{\rm bar}$}

\newcommand{\rr}{${\cal R}$}
\newcommand{\vcirc}{$V_{\rm circ}$}
\newcommand{\mnras}{MNRAS}
\newcommand{\apj}{ApJ}
\newcommand{\apjl}{ApJ}
\newcommand{\aap}{A\&A}
\usepackage{aas_macros}

\title[JD 11.~~A fast bar in NGC~4264] %% give here short title %%
{A MUSE study of the fast bar \\
in the weakly-interacting galaxy NGC~4264}
\author[Virginia Cuomo et al.]{Virginia Cuomo$^{1,*}$,
Enrico Maria Corsini$^{1,2}$, Alfonso J. L. Aguerri$^{3,4}$ \and Victor P. Debattista$^5$}

\affiliation{$^1$Dipartimento di Fisica e Astronomia ”G. Galilei”, Universit\`a di Padova, \\ vicolo dell'Osservatorio 3, I-35122 Padova, Italy \\ $^*$email: {\tt virginia.cuomo@phd.unipd.it} \\[\affilskip]
$^2$INAF - Osservatorio Astronomico di Padova,\\ vicolo dell'Osservatorio 2, I-35122 Padova, Italy \\
$^3$ Instituto de Astrof\'isica de Canarias, \\ calle V\'ia L\'actea s/n, 38205 La Laguna, Tenerife, Spain \\
$^4$ Departamento de Astrof\'isica, Universidad de La Laguna, \\ Avenida Astrof\'isico Francisco S\'anchez s/n, 38206 La Laguna, Tenerife, Spain \\
$^5$ Jeremiah Horrocks Institute, University of Central Lancashire, \\ PR1 2HE Preston, UK

}

\pubyear{2019}
\volume{353}  %% insert here IAU Symposium No.
\jname{Galactic Dynamics in the Era of Large Surveys}
\editors{M. Valluri, \& J. A. Sellwood}
\begin{document}

\maketitle

\begin{abstract}
We present surface photometry and stellar kinematics of NGC~4264, a lenticular galaxy in the region of the Virgo Cluster undergoing a tidal interaction with its neighbour, NGC~4261. We measured the bar radius and strength from SDSS imaging and the bar pattern speed from MUSE integral-field spectroscopy. We find that NGC~4264 hosts a strong and large bar, which is rotating fast. The accurate measurement of the bar rotation rate allows us to exclude that the formation of the bar was triggered by the ongoing interaction.

\end{abstract}

\firstsection % if your document starts with a section,
\section{Introduction}

The bar pattern speed \omegabar\ is the angular velocity of the bar rotating around the galactic centre. It can be parametrised through the bar rotation rate \rr, which is defined as the ratio between the corotation radius \rcor\ and the length of the bar semi-major axis, \rbar. In turn, \rcor\ is measured as the ratio between the circular velocity \vcirc\ of the galaxy and \omegabar. Bars with $1.0 \leq {\cal R} \leq 1.4$ end close to their \rcor\ and are called {\em fast}, while bars with ${\cal R} > 1.4$ are shorter than \rcor\ and are called {\em slow} (\cite[Athanassoula 1992, Debattista \& Sellwood 2000]{Athanassoula1992,Debattista2000}). The value of \omegabar\ is expected to decrease with time because of the angular momentum exchange between the galaxy components and dynamical friction exerted on the bar by the dark matter (DM) halo. A huge amount of DM can efficiently slow down the rotation of the bar, which is expected to be {\em slow} (\cite[Debattista \& Sellwood 1998, Athanassoula et al. 2013]{Debattista1998,Athanassoula2013}). The determination of \rr\ allows both to investigate the secular evolution of barred galaxies and to put constraints on the DM distribution in their central regions.

The only direct way to recover \omegabar\ is to apply the model-independent Tremaine-Weinberg method (\cite[TW, Tremaine \& Weinberg 1984]{Tremaine1984}). It is based on the simple equation $\Omega_{\rm bar}\sin i=\langle V\rangle/\langle X\rangle$, where \omegabar\ is given by the ratio of the luminosity-averaged position $\langle X\rangle$ and line-of-sight (LOS) velocity $\langle V\rangle$ of the stars. These quantities are measured along apertures crossing the bar and located parallel to the disc major axis, when its inclination $i$ is known. The resulting values are fitted with a line and the corresponding slope is proportional to \omegabar. 

Despite its simple formulation, the method has some restrictions: undisturbed morphology and kinematics are required to correctly identify the disc position angle (PA) along which to locate the apertures, so interacting galaxies are generally avoided.

About 100 galaxies have been analysed with the TW method using long-slit (\cite[Corsini 2011]{Corsini2011}) and integral-field spectroscopy (\cite[Debattista \& Williams 2004, Aguerri et al. 2015, Guo et al. 2019, Cuomo et al. 2019a]{Debattista2004,Aguerri2015,Guo2019,Cuomo2019}), hosting only {\em fast} bar. No trends were found between \rr\ and Hubble type or DM content, but so far measurements have large uncertainties ($\langle\Delta\Omega_{\rm bar}/\Omega_{\rm bar}\rangle=0.3, \langle\Delta{\cal R}/{\cal R}\rangle=0.4$).

\section{NGC~4264: an interacting galaxy}

\begin{figure}
    \centering
    \includegraphics[angle=90,scale=0.41]{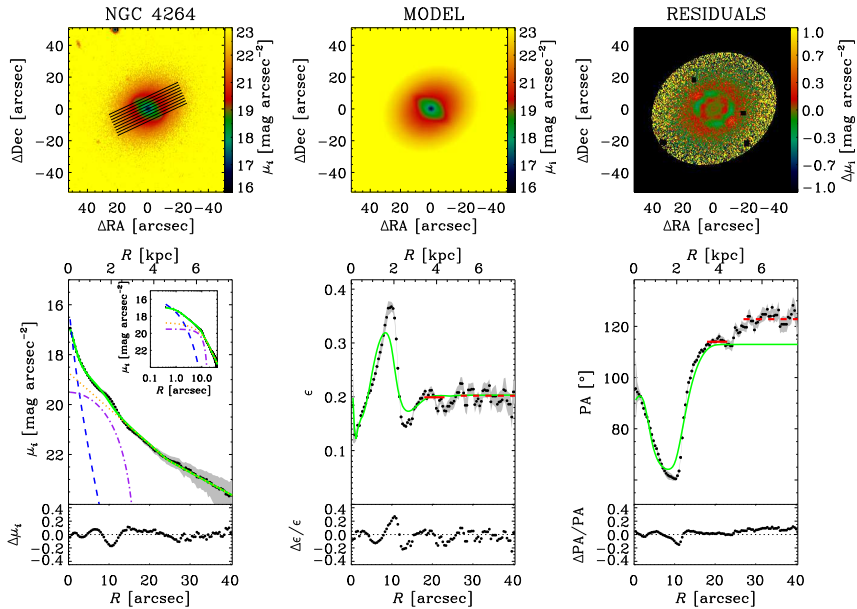}
    \vspace{-0.3cm}
    \caption{Photometric decomposition of NGC~4264. Upper panels: observed, modelled, and residual surface brightness distributions. The FOV is oriented with North up and East left. The apertures are shown as black lines in the observed image. Lower panels: radial profiles of surface brightness, PA, and ellipticity of the observed (black dots with gray error bars) and seeing-convolved modelled image (green solid line) and their difference. The surface brightness radial profiles of the best-fitting bulge (blue dashed line), bar (magenta dash-dotted line), and disc (orange dotted line) are shown for the semi-major axis distance to the galaxy centre. The red solid and dashed segments give the mean values of PA and ellipticity for the inner and outer portions of the disc, respectively.}
    \label{fig:1}
\end{figure}

We analysed the case of NGC~4264, a lenticular galaxy in the region of the Virgo Cluster, with some hints of interaction. In fact, NGC~4264 is located close to the bright elliptical galaxy NGC~4261 at a projected distance of 30 kpc and they are probably gravitationally bound. We measured the surface photometry from the SDSS $i$-band images with the {\sc iraf ellipse} task (\cite[Jedrzejewski 1987]{Jedrzejewski1987}). We observed a twist in the outer disc isophotes. The PA radial profile (Fig.~\ref{fig:1}, bottom-right panel) shows two disc regions characterised by two constant PAs, which differ by $\Delta{\rm PA}\sim10^{\circ}$. This is confirmed by the photometric decomposition we performed using the {\sc gasp2d} code (\cite[M{\'e}ndez-Abreu et al. 2014]{Mendezabreu2014}), in which we included a double-exponential law (\cite[M{\'e}ndez-Abreu et al. 2017]{Mendezabreu2017}) to describe the contribution of the disc to the galaxy surface brightness (Fig.~\ref{fig:1}). 

\section{Characterising the bar}
The correct determination of the disc PA is crucial for a solid measurement of $\Omega_{\rm bar}$ with the TW method (\cite[Debattista 2003]{Debattista2003}). In order to deal with this problem in the case of an interacting galaxy, it is necessary to use the synergy between high-quality photometry and integral-field spectroscopy. The high spatial resolution of MUSE allowed us to define {\em a posteriori} the PA of the disc, and to fine-tune the number and length of the apertures used in the TW analysis.

\begin{figure}[!t]
    \centering
    \includegraphics[scale=0.41]{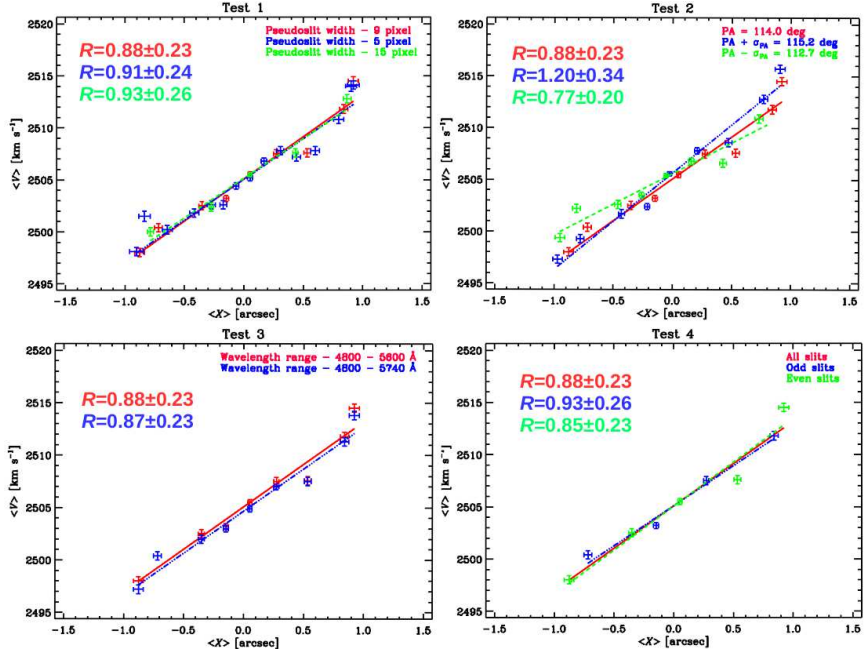}
    \caption{Pattern speed of the bar in NGC~4264 for the different tests performed to assess how the slope of the best-fitting straight line changes as a function of the aperture width (top left panel), PA (top right panel), wavelength range (bottom left panel), and number of apertures (bottom right panel). The reference value in each panel is given by the solid red line.}
    \label{fig:3}
\end{figure}

To characterise the bar it is necessary to measure its radius $a_{\rm bar}$, strength $S_{\rm bar}$, and pattern speed $\Omega_{\rm bar}$. We obtained $a_{\rm bar}=3.2\pm0.5$ kpc by applying three different methods to the SDSS $i$-band image: the photometric decomposition (Fig.~\ref{fig:1}), Fourier analysis, and analysis of the PA of the deprojected isophotes. The Fourier analysis (\cite[Aguerri et al. 2000]{Aguerri2000}) allowed us also to derive $S_{\rm bar}=0.31\pm0.04$. Our results are consistent with typical values found for lenticular galaxies (\cite[Aguerri et al. 2009]{Aguerri2009}). Finally, we measured $\Omega_{\rm bar}=71\pm4$ km s$^{-1}$ kpc$^{-1}$ with the TW method: we defined 9 apertures located parallel to the major axis of the inner disc. To recover the values of $\langle X\rangle$ and $\langle V\rangle$, we collapsed the MUSE datacube between 4800 and 5600 \AA\ in the spectral and spatial directions, respectively. In particular, we extracted the LOS velocity from the resulting spectra. We fitted with the {\sc idl fitexy} task the resulting values (Fig.~\ref{fig:3}, red solid line). 

To recover $V_{\rm circ}=189\pm10$ km s$^{-1}$, we measured the stellar kinematics from the entire MUSE datacube with {\sc ppxf} (\cite[Cappellari \& Emsellem 2004]{Cappellari2004}) and {\sc gandalf} codes (\cite[Sarzi et al. 2006]{Sarzi2006}) on a Voronoi tessellation (\cite[Cappellari, \& Copin 2003]{Cappellari2003}), and we applied the asymmetric drift correction, following the prescription of \cite{Aguerri2015}. Finally, we derived $R_{\rm cor}=2.8\pm0.2$ kpc, and ${\cal R}=0.88\pm0.23$. This means NGC~4264 hosts a {\em fast} bar.

\section{Discussion and conclusions}

We tested the stability of our results by verifying all the ingredients of the TW analysis. We varied width, number, and spectral range used to define the apertures and we found compatible results. On the contrary, changing the PA of the apertures led to different results (Fig.~\ref{fig:3}). In particular, the PA of the outer disc corresponds to ${\cal R}<1.0$, which is unphysical. These tests confirm that the bar of NGC~4264 is living in the inner part of the disc, and that the correct identification of the PA of the disc is crucial for the TW method. All the details are given in \cite{Cuomo2019b}.

This represents a pilot study for further accurate MUSE TW measurements: the values of \omegabar\ and \rr\ for NGC~4264 are amongst the best-constrained ones ever obtained with the TW method ($\Delta\Omega_{\rm bar}/\Omega_{\rm bar} = 0.06$, $\Delta{\cal R}/{\cal R} = 0.26$). 

These results allowed us to constrain the formation mechanism of the bar in NGC~4264. We interpreted the twist of the outer isophotes as suggestive of a warp due to the ongoing interaction with NGC~4261. Since the bar of NGC~4264 is {\em fast}, we concluded that its formation was not triggered by the recent interaction or by a previous interaction with an other galaxy, because this would form a {\em slow} bar (\cite[Martinez-Valpuesta et al. 2017]{MartinezValpueata2017}).

Further accurate measurements of \omegabar\ and \rr\ in a large number of barred galaxies would be of considerable interest to severely test the predictions of numerical simulations about the time evolution of \rbar\ and \omegabar\ as a function of gas content, luminous and DM distribution (\cite[Athanassoula et al. 2013]{Athanassoula2013}).

\end{document}